\documentclass{aastex}
\usepackage{emulateapj5}
\usepackage{graphicx}\usepackage{epsf}
\usepackage{onecolfloat}

\newcommand{\hkpc}{h^{-1}\mathrm{kpc}}

\def\h1{h^{-1}}

\def\LCDM{$\Lambda$CDM}

\def\hMpc{h^{-1}{\ }{\rm Mpc}}

\def\hkpc{h^{-1}{\ }{\rm kpc}}

\begin{document}
\submitted{Submitted to the ApJ 2004 May 21}
\journalinfo{Submitted to the Astrophysical Journal}

\shortauthors{TASITSIOMI, KRAVTSOV, WECHSLER, \& PRIMACK}
\shorttitle{MODELING GALAXY-MASS CORRELATIONS IN SIMULATIONS}
\twocolumn[%
\title{Modeling Galaxy-mass correlations in Dissipationless Simulations}

\author{Argyro Tasitsiomi\altaffilmark{1}, 
  Andrey V. Kravtsov\altaffilmark{1}, 
  Risa H. Wechsler\altaffilmark{1,2},
  Joel R. Primack\altaffilmark{3}}
\vspace{2mm}
\begin{abstract}  
  We use high-resolution dissipationless simulations of the
  concordance flat $\Lambda$CDM model to make predictions for the
  galaxy--mass correlations and compare them to the recent SDSS weak
  lensing measurements of \cite{sheldon_etal04}.  The simulations
  resolve both isolated galaxy-size host halos and satellite halos
  (subhalos) within them.  We use a simple scheme based on the
  matching of the circular velocity function of halos to the galaxy
  luminosity function and on using the observed density-color
  correlation of the SDSS galaxies to assign luminosities and colors
  to the halos.  This allows us to closely match the selection
  criteria used to define observational samples.  The simulations
  reproduce the observed galaxy--mass correlation function and the
  observed dependence of its shape and amplitude on luminosity and
  color, if a reasonable amount of scatter between galaxy luminosity
  and circular velocity is assumed. We find that the luminosity
  dependence of the correlation function is primarily determined by
  the changing relative contribution of central and satellite galaxies
  at different luminosities. The color dependence of the galaxy--mass
  correlations reflects the difference in the typical environments of
  blue and red galaxies. We compare the cross-bias, $b_x\equiv b/r$,
  measured in simulations and observations and find a good agreement
  at all probed scales.  We show that the galaxy--mass correlation
  coefficient, $r$, is close to unity on scales $\gtrsim 1 \hMpc$.
  This indicates that the cross bias measured in weak lensing
  observations should measure the actual bias $b$ of galaxy clustering
  on these scales.  In agreement with previous studies, we find that
  the aperture mass-to-light ratio is independent of galaxy color in
  the range of luminosities probed by observational samples. The
  aperture mass scales approximately linearly with luminosity at $L_r>
  10^{10}h^{-2}\ \rm L_{\odot}$, while at lower luminosities the
  scaling is shallower: $M_{\Delta\Sigma}\propto L_r^{0.5}$.  We show
  that most of the luminous galaxies ($M_r<-21$) are near the centers
  of their halos and their galaxy--mass correlation function at
  $r\lesssim 100h^{-1}\ \rm kpc$ can therefore be interpreted as the
  average dark matter density profile of these galaxies. Finally, we
  find that for galaxies in a given narrow luminosity range, there is
  a broad and possibly non-gaussian distribution of halo virial
  masses. Therefore, the average relation between mass and luminosity
  derived from the weak lensing analyses should be interpreted with
  caution.

\end{abstract}


]

\altaffiltext{1}{Dept. of Astronomy and Astrophysics,
       Kavli Institute for Cosmological Physics,
       The University of Chicago, Chicago, IL 60637;
       {\tt iro@oddjob.uchicago.edu, andrey@oddjob.uchicago.edu, risa@cfcp.uchicago.edu}}
\altaffiltext{2}{Hubble Fellow, Enrico Fermi Fellow}
\altaffiltext{3}{Physics Department, University of California, Santa Cruz, CA 95064
       {\tt joel@scipp.ucsc.edu}}

\section{Introduction}
\label{sec:intro}

Understanding the processes that shape the clustering of dark matter
and galaxies is one of the main goals of observational cosmology.
Modern large redshift surveys, such as the Two Degree Field Galaxy
Redshift Survey ~\citep[2dFGRS,][]{colless_etal01} and the Sloan
Digital Sky Survey ~\citep[SDSS,][]{york_etal00}, allow measurements
of galaxy and galaxy--mass correlations and of their dependence on
galaxy properties and environment with unprecedented accuracy.

Concurrently, cosmological $N$-body simulations have developed
into a powerful tool for calculating the gravitational clustering of
collisionless dark matter in hierarchical cosmologies with
well-specified initial conditions. The main obstacle in direct
comparisons between models and data is understanding the dependence of
theoretical predictions on both the relatively straightforward physics
of gravitational clustering and the more complex physics of galaxy
formation. The former determines the distribution of dark matter,
while the latter affects the relationship between the distribution of
galaxies and mass, the ``bias''. The complexity of processes operating
during galaxy formation on a very wide range of scales makes it
difficult to include them directly in simulations, although efforts in
this direction are ongoing \citep[e.g.,][]{katz_etal99, white_etal01,
yoshikawa_etal01, pearce_etal01, berlind_etal03,weinberg_etal04}.

Given the difficulty of large-scale galaxy formation simulations, two
simpler approaches have recently been pursued to make theoretical
predictions for galaxy properties.  In the first, the galaxies are
identified with dark matter halos and subhalos in dissipationless
cosmological simulations \citep[e.g.,][]{colin_etal99,
kravtsov_klypin99, neyrinck_etal03,kravtsov_etal04}. In the second
hybrid approach, collisionless $N$-body simulations are combined with
a semi-analytic treatment of galaxy formation
\citep[e.g.,][]{kauffmann_etal97, governato_etal98, kauffmann_etal99a,
kauffmann_etal99b, kolatt_etal99, benson_etal00a, benson_etal00b,
somerville_etal01, wechsler_etal01, berlind_etal03}.

Galaxy-galaxy lensing is a relatively new but important observational
probe of the relation between galaxies and dark matter. It directly
measures the galaxy--mass cross-correlation function around galaxies
of different types and environments \citep[see,
e.g.,][]{brainerd_etal96, fischer_etal00, hoekstra_etal01,
hoekstra_etal04, mckay_etal01, sheldon_etal01,
sheldon_etal04,smith_etal01,wilson_etal01}.

Several theoretical studies of the galaxy--mass correlations in
cosmological simulations have been carried out in the last three years
\citep{guzik_seljak01,guzik_seljak02,yang_etal03,weinberg_etal04}.
Still, there appear to be several discrepancies between the various
model results and observations, which are yet to be properly
understood.  \cite{guzik_seljak01} used intermediate-resolution dark
matter only simulations combined with a semi-analytic model to
identify galaxies and make predictions for the galaxy--mass
correlation function.  They compared these results to the first weak
lensing detection from SDSS \citep{fischer_etal00}, and found the
amplitude of the galaxy--mass correlation in their model to be
systematically higher.  They attributed this discrepancy to the
differences between the luminosity function used in their
semi-analytic prescription and the observed luminosity function.
\cite{yang_etal03}, using the same theoretical model
(intermediate-resolution dark matter simulations combined with the
\citealt{kauffmann_etal99a} semi-analytic galaxy formation prescription),
compared to more extensive observational results from
\cite{mckay_etal01}.  They also found that the galaxy--mass
correlation function in the simulations was systematically higher than
the observational measurements, by about a factor of two.
Consequently, they found mass-to-light ratios about a factor of two
higher than those observed.  The study of \cite{weinberg_etal04}
investigated several of the same statistics, based on galaxies
identified in SPH simulations.  They found that for systems with small
baryonic masses, their dark matter-to-baryon ratios agreed with the
mass-to-light ratios, derived from weak lensing data, presented by
\cite{mckay_etal01}.  For large masses, however, the baryonic masses
of their simulated galaxies were high by a factor of 1.5-2 compared to
what is required for agreement with the SDSS data.  In each of these
studies, discrepancies with the data may have been due to
discrepancies of the model luminosity functions with that observed.
\cite{guzik_seljak02} used a more phenomenological approach, and
developed a formalism in the context of the halo model to fit the
\cite{mckay_etal01} data. They extracted information on, e.g., the
dependence of the virial mass-to-light ratio on luminosity, the
typical mass for galaxies, and the fraction of galaxies in groups and
clusters.

The most recent observational study of galaxy--galaxy lensing by the
SDSS collaboration \citep[][hereafter S04]{sheldon_etal04}, has
significantly improved the accuracy of galaxy correlation
measurements, and may shed light on many of the previous uncertainties.  The
spectroscopic sample of 127,001 lensing galaxies was combined with
more than $9 \times 10^{6}$ source galaxies with photometric
redshifts, which allowed a detailed study of the relation between mass
and light for several luminosity bands, morphological types, and on
scales from 20 $\hkpc$ to 10 $\hMpc$ (physical). In addition to
verifying the findings of previous studies, S04 identified new
features in the data, including a scale-dependent luminosity (cross) bias.

Besides being a general test of the Cold Dark Matter paradigm,
comparisons with cosmological simulations can be used to gain a deeper
insight into the interpretation of these new observational results.
Furthermore, they can shed light on to what extent weak lensing
results can be used to estimate halo masses, or to learn about the
average dark matter halo density profiles. In this paper we use
high-resolution dissipationless simulations of the concordance
$\Lambda$CDM model to study the galaxy--mass correlation function with
specific emphasis on comparing to the observational measurements of
S04. Results of several recent studies suggest that gravitational
dynamics is the dominant mechanism shaping galaxy clustering, at least
in the simple case of galaxies selected above a luminosity or mass
threshold \citep{kravtsov_klypin99, kravtsov_etal04,zentner_etal04}.
\citet{kravtsov_etal04}, using dark-matter only simulations which
resolve galactic mass subhalos, matched galaxies of a given luminosity
to a population of halos and subhalos of a given circular velocity and
the same number density, and found excellent agreement with the
galaxy-galaxy correlation functions measured in the SDSS. This study
extends that approach to investigate the galaxy--mass correlations,
and makes detailed comparisons to the new \citet{sheldon_etal04}
results.

The paper is organized as follows. In \S~\ref{sec:sim} and
\S~\ref{sec:haloid} we describe the simulations and the halo
identification algorithm. The halo samples used in our analysis are
described in \S~\ref{sec:samples}. The main results, including a
detailed comparison with the most recent SDSS measurements, are
presented in \S~\ref{sec:results}.  In \S~\ref{sec:discussion} and
\S~\ref{sec:conclusions} we discuss and summarize our results and
conclusions.

\begin{table}[tb]
\label{tab:sim}
\caption{Simulation parameters}
\begin{center}
\small
\begin{tabular}{cccccc}
\tableline\tableline\\
\multicolumn{1}{c}{Name}&
\multicolumn{1}{c}{$\sigma_8$}&
\multicolumn{1}{c}{$L_{\rm box}$}  &
\multicolumn{1}{c}{$N_{\rm p}$}  &
\multicolumn{1}{c}{$m_{\rm p}$}  &
\multicolumn{1}{c}{$h_{\rm peak}$} 
\\
\multicolumn{1}{c}{}&
\multicolumn{1}{c}{}&
\multicolumn{1}{c}{$h^{-1}\rm Mpc$} &
\multicolumn{1}{c}{} &
\multicolumn{1}{c}{$h^{-1}\rm\ M_{\odot}$} & 
\multicolumn{1}{c}{$h^{-1}\rm\ kpc$}  
\\
\\
\tableline
\\
$\Lambda$CDM$_{\rm 60a}$ & 1.0  & 60  & $256^3$ & $1.07\times 10^9$ & $1.9$\\
$\Lambda$CDM$_{\rm 60b}$ & 0.9  & 60  & $256^3$ & $1.07\times 10^9$ & $1.9$\\
$\Lambda$CDM$_{\rm 60c}$ & 0.9  & 60  & $256^3$ & $1.07\times 10^9$ & $1.9$\\
$\Lambda$CDM$_{\rm 80a}$ & 0.75 & 80  & $512^3$ & $3.16\times 10^8$ & $1.2$\\
$\Lambda$CDM$_{\rm 80b}$ & 0.9  & 80  & $512^3$ & $3.16\times 10^8$ & $1.2$\\
$\Lambda$CDM$_{\rm 120}$ & 0.9  & 120  & $512^3$ & $1.07\times 10^9$ & $1.8$\\
\\
\tableline
\end{tabular}
\end{center}
\label{tab:simparam}
\end{table}

\section{Simulations}
\label{sec:sim}

For each simulation in this study we assume the concordance flat
{\LCDM} model: $\Omega_0=1-\Omega_{\Lambda}=0.3$, $h=0.7$, where
$\Omega_0$ and $\Omega_{\Lambda}$ are the present-day matter and
vacuum densities, and $h$ is the dimensionless Hubble constant defined
as $H_0\equiv 100h{\ }{\rm km\ s^{-1}\,Mpc^{-1}}$.  This model is
consistent with recent observational constraints
\citep[e.g.,][]{spergel_etal03,tegmark_etal04}. To study the effects
of the power spectrum normalization, the effects of box size, and
cosmic variance, we consider a set of simulations listed in
Table~\ref{tab:simparam}. The simulations are labeled according to
their box size. 

Three different simulations of $60\hMpc\approx
85.71$~Mpc box with $256^3\approx 1.67\times 10^7$ particles are used.
One of these three simulations is normalized to $\sigma_8=1.0$, where
$\sigma_8$ is the rms fluctuation in spheres of $8h^{-1}{\ }{\rm Mpc}$
comoving radius.  This simulation was used previously to study the
halo clustering and bias by \citet{kravtsov_klypin99} and
\citet{colin_etal99}; the reader is referred to these papers for
further details.  The other two $60\hMpc$ simulations have a lower
normalization of $\sigma_{8}=0.9$, and differ only in the realization
of the Gaussian initial conditions.  The second subset of simulations
followed the evolution of $512^3\approx 1.34\times 10^8$ particles in
a $80\hMpc\approx 114.29$~Mpc box. These simulations have different
power spectrum normalizations of $\sigma_8=0.75$ and $\sigma_8=0.9$,
but were started from the same realization of the initial conditions.
We also use a simulation of a 120 $\hMpc$ box with $512^3$ particles
and $\sigma_{8}=0.9$.

The simulations were run using the Adaptive Refinement Tree $N$-body
code \citep[ART;][]{kravtsov_etal97,kravtsov99}. The ART code reaches
high force resolution by refining all high-density regions with an
automated refinement algorithm. The criterion for refinement is the
mass of particles per cell. In the $\Lambda$CDM$_{\rm 60a}$ the code
starts with zeroth level uniform $512^3$ grid and refines an
individual cell only if the mass exceeds $n_{\rm th}=5$ particles,
independent of the refinement level. In terms of overdensity, this
means that all regions with overdensity higher than $\delta = n_{\rm
th}{\ }2^{3L}/\bar{n}$, where $\bar{n}$ is the average number density
of particles in the cube, are refined to the refinement level
$L$. Thus, for the $\Lambda$CDM$_{\rm 60a}$ simulation , $\bar{n}$ is
$1/8$.  The peak formal dynamic range reached by the code in this
simulation is $32,768$, which corresponds to a peak formal resolution
(the smallest grid cell) of $h_{\rm peak}=1.83\hkpc$; the actual force
resolution is $\approx 2h_{\rm peak}=3.7\hkpc$
\citep[see][]{kravtsov_etal97}.  In the higher-resolution
$\Lambda$CDM$_{80}$ simulations the initial grid is $256^3$ and the
refinement criterion is level- and time-dependent. At the early stages
of evolution ($a<0.65$) the thresholds are set to 2, 3, and 4 particle
masses for the zeroth, first, and second and higher levels,
respectively. At low redshifts, $a>0.65$, the thresholds for these
refinement levels are set to 6, 5, and 5 particle masses.  The lower
thresholds at high redshifts are set to ensure that collapse of
small-mass halos is followed with higher resolution. The maximum
achieved level of refinement is $L_{\rm max}=8$, which corresponds to
a comoving cell size of $1.22\hkpc$.  As a function of redshift the
maximum level of refinement is equal to $L_{\rm max}=6$ for $5<z<7$,
$L_{\rm max}= 7$ for $1<z<5$, $L_{\rm max}\geq 8$ for $z<1$. The peak
formal resolution is $h_{\rm peak}\leq 1.2\hkpc$ (physical). The
refinement criteria for the $\Lambda$CDM$_{120}$ simulation are
similar to those of the $\Lambda$CDM$_{80}$ runs, except that
initially the entire volume is resolved with a $1024^3$ grid. The
parameters of the simulations are summarized in
Table~\ref{tab:simparam}.

\section{Halo identification}
\label{sec:haloid}

\subsection{The algorithm}

A variant of the Bound Density Maxima halo finding algorithm
\citep{klypin_etal99} is used to identify halos and the subhalos
within them.  The details of the algorithm and parameters used in the
halo finder can be found in \citet{kravtsov_etal04}. The main steps of
the algorithm are identification of local density peaks (potential
halo centers) and analysis of the density distribution and velocities
of the surrounding particles to test whether a given peak corresponds
to a gravitationally bound clump. More specifically, we construct
density, circular velocity, and velocity dispersion profiles around
each center and iteratively remove unbound particles using the
procedure outlined in \citet{klypin_etal99}.  We then construct final
profiles using only bound particles and use these profiles to
calculate properties of halos, such as the circular velocity profile $V_{\rm
  circ}(r)=\sqrt{GM(<r)/r}$ and compute the maximum circular velocity
$V_{\rm max}$.  Our completeness limit is approximately $50$
particles. This limit corresponds to the mass below which cumulative
mass and velocity functions start to flatten significantly
\citep{kravtsov_etal04}. In the following analysis, only halos with
masses $M>50m_{\rm p}$ are considered.

\subsection{Halo Classification: host halos, satellites, and central galaxies}

In this study we distinguish between {\it host} halos with centers
that do not lie within any larger virialized system and {\it subhalos}
located within the virial radii of larger systems. Below the 
terms {\em satellites}, {\em subhalos}, and {\em substructure} are used 
interchangeably.  To classify the halos, we calculate the formal
boundary of each object as the radius corresponding to an enclosed
overdensity of 180 with respect to the mean density of the universe. 
Halos whose centers are located within the boundary of a larger
mass halo are classified as {\it subhalos} or {\it satellites}.

The halos that are not classified as satellites are identified as {\it
host} halos. Note that the center of a host halo is not considered to
be a subhalo. Thus, host halos may or may not contain any subhalos
with circular velocity above the threshold of a given sample. The host
centers, however, are included in clustering statistics because we
assume that each host harbors a {\it central} galaxy at its center.
Therefore, the total sample of galactic halos contains central and
satellite galaxies. The former have the positions and maximum circular
velocities of their host halos, while the latter have the positions and
maximum circular velocities  of subhalos.

In the observed universe, the analogy is simple. In a cluster, for
example, the brightest central galaxy that typically resides near the
center would be associated with the cluster host halo in our
terminology. All other galaxies within the virial radius of the
cluster would be considered ``satellites'' associated with subhalos.

\section{The Galaxy Samples}
\label{sec:samples}
In this study, our galaxy sample is created by assigning realistic
SDSS luminosities and colors to dark matter halos.  To construct mock
galaxy catalogs for comparison with observations, one must define
selection criteria for particular halo properties to mimic the
selection function of the observational sample as closely as possible.
Halo mass is often used to define halo catalogs: e.g., a catalog can
be constructed by selecting all halos in a given mass range.  However,
the mass and radius are poorly defined for the satellite halos due to
tidal stripping --- which alters a halo's mass and physical extent
\citep[see][]{klypin_etal99}.  Therefore, we use the maximum circular
velocity, $V_{\rm max}$, as a proxy for the halo mass.  For isolated
halos, $V_{\rm max}$ and the halo's virial mass are directly related.
For subhalos, $V_{\rm max}$ will experience secular decrease but at a
relatively slow rate \citep{kravtsov_etal04b}.

To mimic the observational selection function, $r$-band luminosities
are assigned to the halos as follows. We match the cumulative velocity
function $n(>V_{\rm max})$ of the halos to the SDSS observed $r$-band
cumulative luminosity function \citep{blanton_etal03}.  Note that
$n(>V_{\rm max})$ includes both isolated host halos and subhalos.  We
use the $r$-band data since it is in that band that the SDSS spectra
were selected, in which the luminosity function is more reliably
measured observationally, and on which most SDSS analyses focus.  The
mean redshift of SDSS galaxies in the main sample is $z=0.1$, and this
is also quite close to the weighted mean for the lens galaxies used by
S04.  We thus start by matching the $M_r(z=0.1)$ luminosity function
to the halos identified at $z=0.1$, obtaining the average $V_{\rm max}
-M_{r}$ relation by matching $n(>V_{\rm max})$ to $n(<M_{\rm r})$.
This was the same method used to assign luminosities to halos in
\citet{kravtsov_etal04}, in which galaxy clustering properties were
reproduced remarkably well.  To match what was done with the S04
magnitudes, we $K$-correct the $z=0.1$ magnitudes to $z=0$ (using the
code {\tt kcorrect v3\_2}, described in \citealt{blanton_etal03c}).

Note that the magnitudes in S01 have not been corrected for evolution
effects, and we use the \citet{blanton_etal03} luminosity function
appropriate for $z=0.1$.  The brightest galaxies in the S01 sample are
typically at a somewhat higher redshift. Therefore, if we had included
evolution effects the galaxy-mass correlation of our brightest sample
would be lower than is shown below, since the brightest objects would
then be too bright to be included in the corresponding S01 sample.
However, the effect is quite small in the magnitude range we consider:
smaller than the uncertainty due to cosmic variance and luminosity
assignment for galaxies. The evolution correction for the brightest
galaxies is less than 0.1 mags, and would result in a less
than 10\% correction to the $\Delta\Sigma$ amplitude.

\begin{figure}[t]
\centerline{\epsfxsize=3.5truein\epsffile{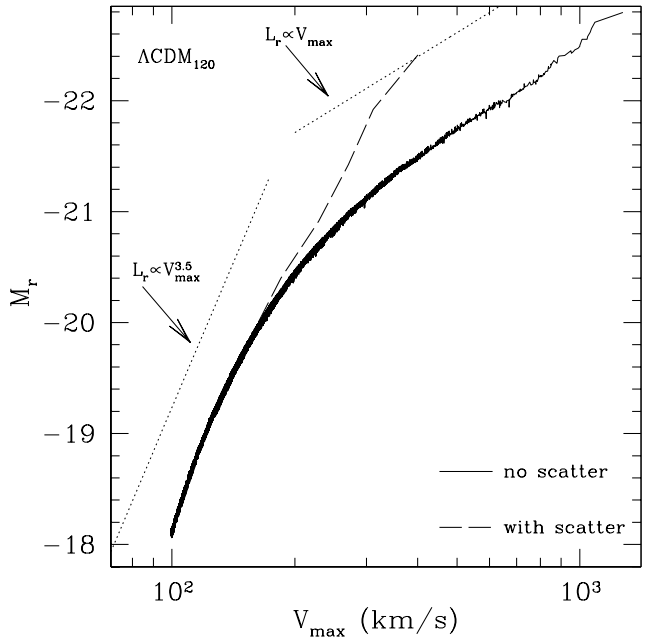}}
\caption{The $r$-band absolute magnitude (the magnitudes are $M-5\log h$) 
  vs. the maximum circular velocity of halos. The solid curve is
  obtained by matching the cumulative velocity function $n(>V_{\rm
    max})$ to the SDSS luminosity function $n(<M_{r})$, both at
  $z=0.1$. The dashed line denotes the mean $V_{\rm max}$ as a function
of $M_r$, after scatter between luminosity and $V_{\rm
    max}$ is introduced as described in the text.  For comparison, the dotted lines show a power
  law $L_{r} \propto V_{\rm max}^{\alpha}$, with $\alpha=3.5$ and
  $\alpha=1$.
\label{fig:vmax_lum}}
\end{figure}

Observations indicate that there is substantial scatter in the
relation around the mean Tully-Fisher (TF) and Faber-Jackson
correlations for spiral and elliptical galaxies.  The magnitude of the
scatter varies for different galaxy samples, wavebands, slopes of the
correlation, the corrections applied, etc.  \citet{willick_etal97}
found that the TF residuals are consistent with a Gaussian
distribution with the standard deviation equal to the rms scatter
after various corrections to the absolute magnitudes are applied.  The
{\sl rms} values for this scatter in magnitude at a given velocity in
red and infrared wavelengths range from $\sim 0.2$ to $\sim 0.9$
magnitudes \cite[e.g.,][]{aaronson_mould83, willick_etal97,
  verheijen01}.

Motivated by these findings, we introduce scatter in the relation
between $V_{\rm max}$ and $M_r$, assuming that at a given $V_{max}$
there is a Gaussian distribution of possible $M_{r}$ values.  The
luminosities are assigned to halos using an iterative procedure that
preserves the observed luminosity function.  We start by deriving the
mean $M_{r}-V_{\rm max}$ relation by matching the cumulative velocity
function in the simulation with the observed luminosity function.
Given our assumption of a Gaussian scatter in $M_{r}$ with fixed
standard deviation $\sigma_{M_{r}}$ or, equivalently, in natural log
luminosity with a fixed standard deviation $\sigma_{\ln{L}}$, the
scatter in velocity at fixed luminosity will be given by
$\sigma_{\ln{V_{max}}} = d{\rm log}V_{max}/d{\rm log}L \times
\sigma_{\ln{L}}$, where the $V_{max}-L$ relation is taken to be
locally well approximated by a power law. As we cannot change the
observed galaxy luminosities and luminosity function, to assign a
luminosity accounting for scatter we choose to perturb the values of
halo $V_{\rm max}$ using the expression above and using the mean
$V_{\rm max}-M_r$ relation.  Note that the maximum circular velocities
are perturbed only during the procedure of assigning luminosities to
halos and the actual $V_{max}$ of an object does not change.
In each iteration, we thus perturb the velocities by a Gaussian in
natural log with this standard deviation, construct a new circular
velocity function, and match this to the observed luminosity function
to assign new luminosities to halos.  The above procedure gives a new mean
$M_{r}-V_{\rm max}$ relation which is used to assign scatter for the
next iteration. The iterations are repeated until convergence.

In the analyses presented below we assume a standard deviation of
1.5 magnitudes for the $M_{r}$ distribution at fixed $V_{\rm max}$.
This magnitude of scatter is needed to match the amplitude of the
galaxy-mass correlation for the bright samples with the observed
values. Although this {\sl rms} is larger than the values derived from
the TF analyses, we note that the latter are derived for the
luminosities corrected for the effects of extinction, inclination,
etc. We, on the other hand, are applying scatter to the raw
luminosities uncorrected for any such effects. For example, the
correction due to internal extinction alone can be as high as one
magnitude \citep[e.g., ][]{verheijen01}.  We make the simplifying
assumption that the scatter is just a function of $V_{max}$, and do not
model it as a function of type for a given $V_{max}$.  Including all
galaxy types in a sample is also likely to increase the scatter
\citep{kannappan_etal02, courteau_etal04}.  The magnitude of scatter
we assume is therefore perfectly consistent with the
observations. Given that the scatter does affect the amplitude of the
galaxy-galaxy and galaxy-mass correlation for bright galaxies, it
should not be neglected.  However, since the fiducial value we assume
is still uncertain, we note where our results may be sensitive to this
assumption.

The relation between the $r$-band absolute magnitude and the maximum
halo circular velocity obtained with and without accounting for
scatter is shown in Figure~\ref{fig:vmax_lum}. The scatter has a
significant effect. Without the scatter, the relation has a running
slope. Including the scatter transforms the relation into almost a
power law $L_{r} \propto V_{\rm max}^{3.5}$.  The effect of scatter is
similar to the Malmquist bias: because the abundance of galaxies
increases with decreasing luminosity, more low luminosity objects
scatter into sample at a fixed mass than high luminosity objects.
As we show below, the effect of scatter on the amplitude of $\Delta\Sigma$
is strong at high luminosities and is small at $L\lesssim L_{\ast}$.

The limited size of the simulation box puts an upper limit on the
luminosities of galaxies that can be studied reliably, and the
completeness limit of our halo catalogs imposes a lower limit on the
luminosities of objects that can be considered.  Thus, we will not
present a comparison of our results to the S04 sample as a whole,
since our sample lacks both the faintest of their objects ($-18 \le
M_{r} \le -17$), and, most importantly, their brightest bin ($-24.0
\le M_{r} \le -22.2$).  To obtain reliable statistics for objects that
bright in simulations, a considerably larger box size is needed.
Instead, we try to mimic the intermediate luminosity samples of S04 as
closely as possible.

To investigate trends in other bands and with color, magnitudes are
assigned to subhalos in the remaining four SDSS bands ($u$, $g$, $i$,
and $z$) using the observed relation between local galaxy density and
color (similar to the procedure described in \citealt{wechsler03}, 
\citealt{wechsler_etal04}).  We first select a volume-limited
sample of SDSS galaxies from the first public data release
\citep[DR1,][]{abazajian_etal03}.  We use the CMU-Pitt Value Added
Catalog to compile a subsample of these galaxies which is not
sensitive to edge effects, and to get a local density measurement for
each galaxy: the projected distance to the tenth nearest galaxy
neighbor, brighter than $M_r -5\log{h} = -19.7$ and within $cz = 1000
\ \rm{km/s}$.  We measure the identical quantity for our mock
galaxies.  For each mock galaxy, we then choose a real SDSS galaxy
which has a similar $r$-band luminosity and nearest neighbor distance,
and assign the colors of this galaxy to the mock galaxy (bins are
chosen to contain $\sim 50$ galaxies).  Assigning colors in this way
allows us to test whether the color- and band- dependent trends in the
data can be reproduced with dark matter halos, under the simple
assumption that the dominant variables for determining galaxy colors
are its luminosity (or the circular velocity of its subhalo) and the
local environment (as is indicated by \citealt{blanton_etal03b}).

\begin{figure}[tb]
\vspace{-0.5cm}
\centerline{\epsfxsize=4truein\epsffile{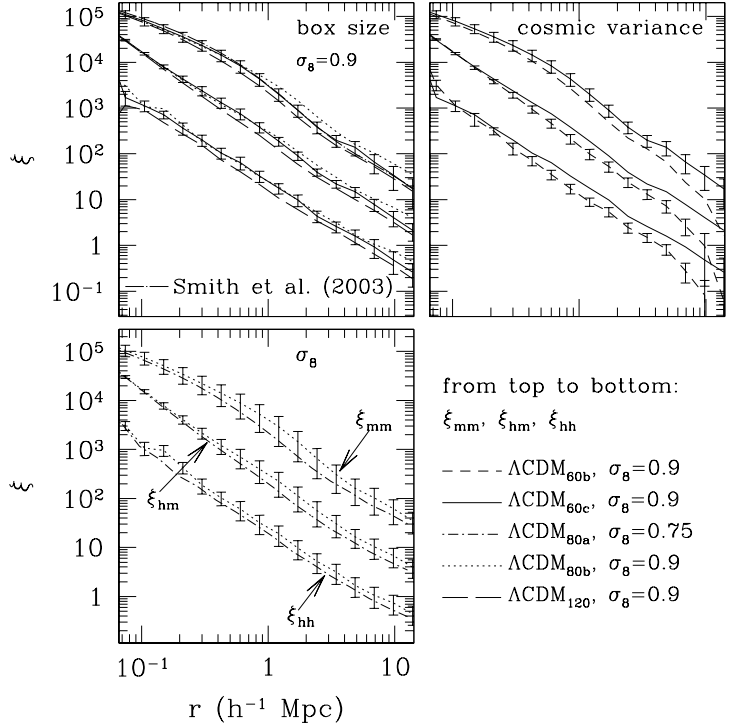}}
\caption{{\em Top left panel:} The effect of the box size. {\em Top
  right panel:} The effect of cosmic variance.  {\em Bottom panel:}
  the effect of power spectrum normalization.  In each panel we plot,
  from top to bottom, the {\it mass-mass} ($\xi_{\rm mm}$), the {\it
  halo--mass} ($\xi_{\rm hm}$), and the {\it halo--halo} ($\xi_{\rm
  hh}$) two-point correlation functions.  For all boxes, halos
  (including subhalos) of the same number density are used to
  calculate the correlation functions. For clarity purposes, the halo--mass and
  the mass-mass functions are scaled up by a factor of 10 and 100,
  respectively. The jackknife resampling error bars are plotted. The
  parameters of the simulations are summarized in
  Table~\ref{tab:simparam}. All correlation functions are presented at
  $z=0.1$.
\label{fig:boxes}}
\end{figure}

\section{Error estimates}

In what follows, all distances, both three-dimensional and projected,
are comoving. We study the mass-mass and galaxy--mass two-point
correlation functions at distances from $20 \hkpc$ up to approximately
$15 \hMpc$, except for the galaxy--galaxy correlation function, which
cannot be measured down to $20h^{-1}$~kpc in our runs.  The simulation
data is divided into 20 equal logarithmic bins in $r$. As an estimate
of the error in the correlation function in each bin we use the
maximum of the Poisson error and the jackknife resampling error. The
latter is an estimate of the cosmic variance and is calculated by
dividing the simulation box in eight sub-volumes. The jackknife
variance $\sigma^{2}_{\rm JK}$ of the quantity $\chi$ is then

\begin{equation}
\sigma^{2}_{\rm JK}= \frac{N-1}{N} \sum_{i=1}^{N}
(\chi_{i} -\bar{\chi_{i}})^{2}.
\end{equation}
Here, $N=8$ is the number of sub-volumes, $\chi_{i}$ is
the mean of $\chi$ when the $i$-th sub-volume is excluded, and
$\bar{\chi_{i}}$ is the mean of $\chi_{i}$. Typically, the jackknife
resampling error is larger than the Poisson error at all scales.

\section{Cosmic variance and the effects of box size and $\sigma_8$ }
\label{sec:boxes}
In this section, results for the different simulations are compared to
understand the effects of the box size, of different power spectrum
normalizations ($\sigma_{8}$), and of cosmic variance on the two-point
correlation functions.  The results are shown in
Figure~\ref{fig:boxes}. In each panel the mass--mass, halo--mass, and
halo--halo correlations are plotted from top to bottom, respectively.
Each is measured for a volume-limited sample that consists of objects
of the same number density for all boxes.  The halo--mass and
mass--mass correlation functions are displaced by a factor of 10 and
100 upward, respectively, for clarity.

\begin{figure*}[th]
\centerline{\epsfxsize3.5truein\epsffile{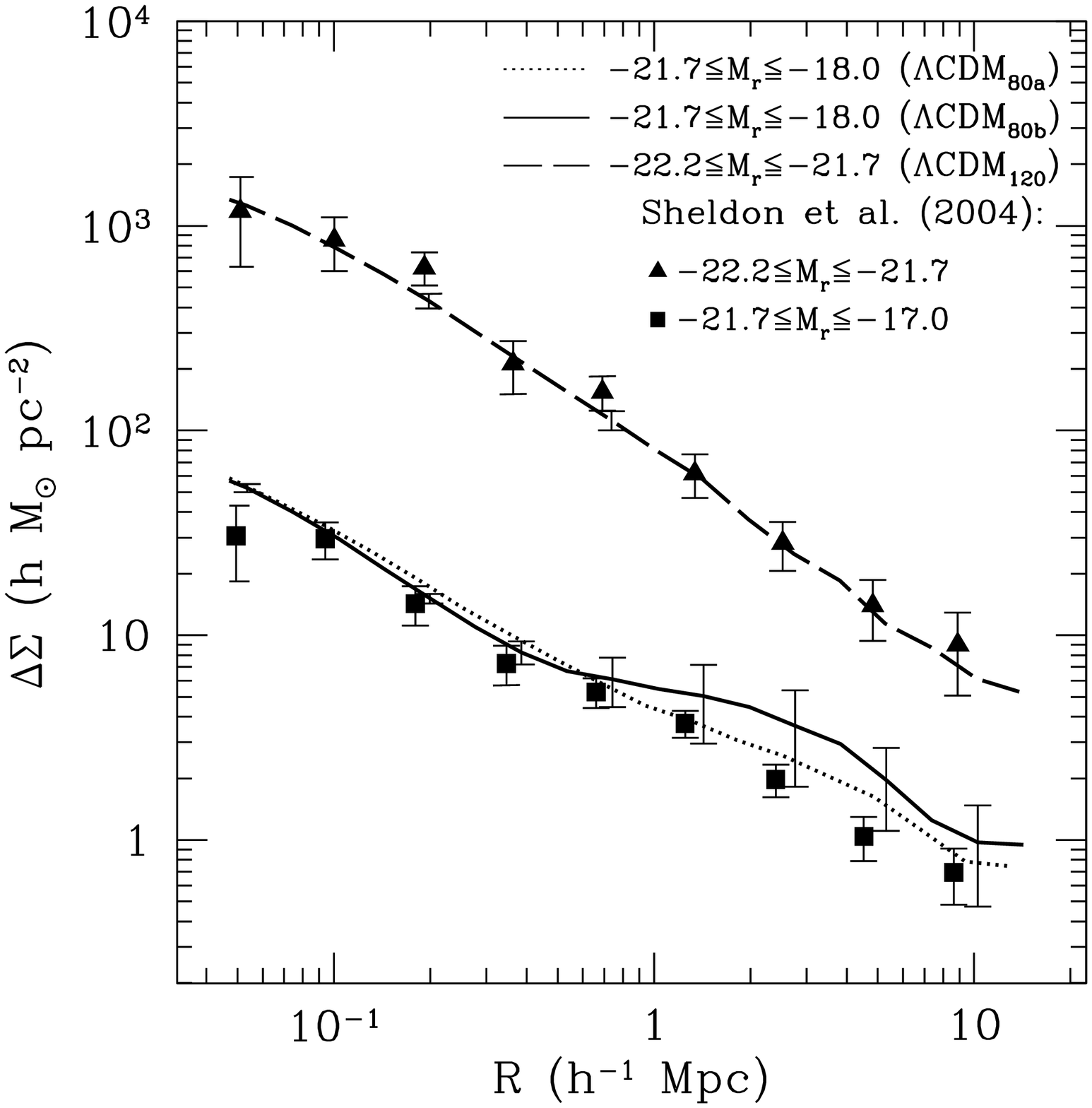}\hspace{0.5cm}\epsfxsize3.5truein\epsffile{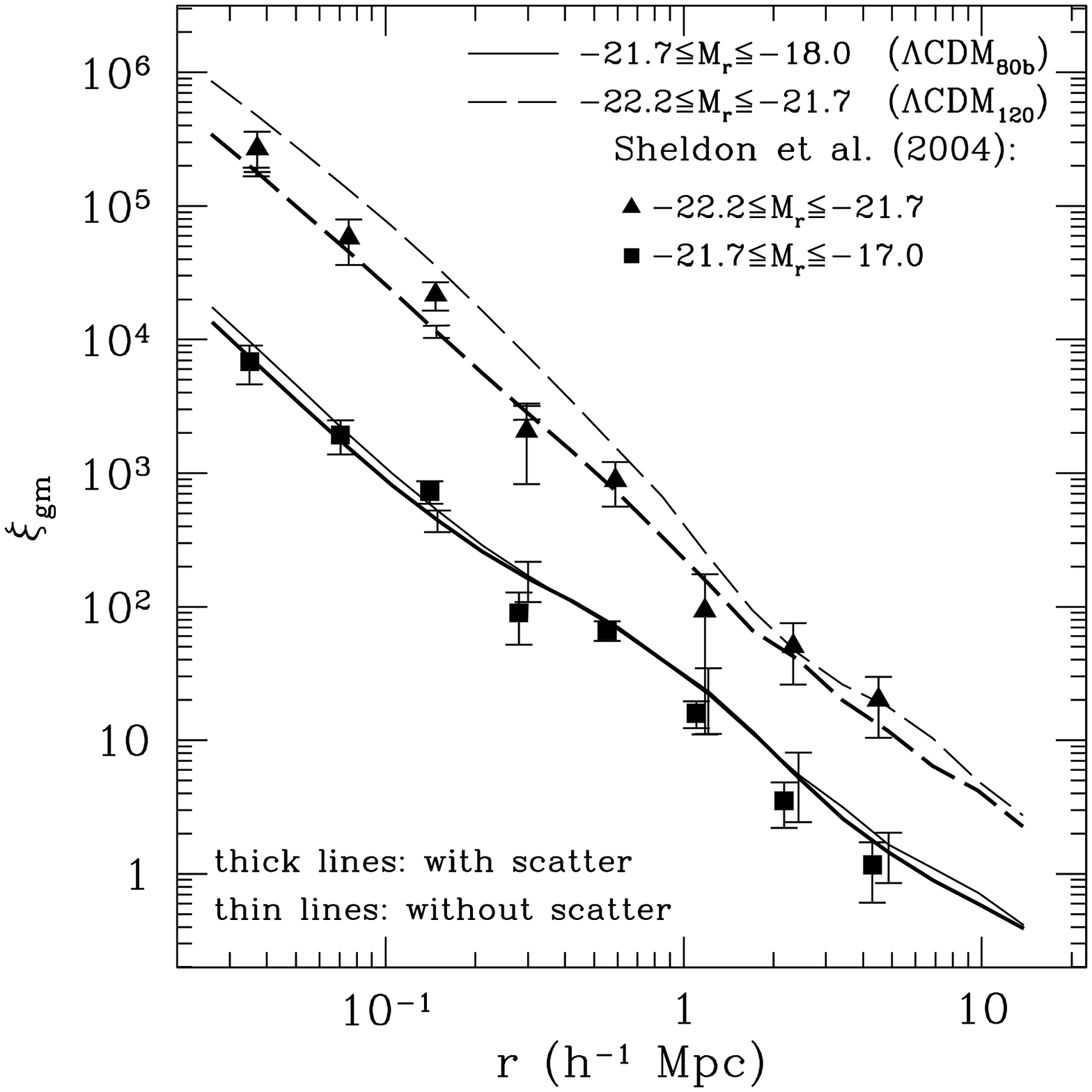}}
\caption{{\it Left panel:} $\Delta \Sigma$ as a function of 
  projected separation $R$ as measured by \cite{sheldon_etal04} ({\it
    points}) and as calculated from the simulations ({\it lines}).
  The results are shown for two $r$-band luminosity bins: $-22.2\le
  M_{r}\le -21.7$ ({\it triangles/dashed line}) and $-21.7\le M_{r}\le
  -17.0 (-18.0)$ ({\it squares} and {\it solid line}, respectively).
  The effect of $\sigma_8$ can be seen for the fainter sample for
  which we plot the results for the $\sigma_{8}=0.75$ ({\it dotted}
  line) and $\sigma_{8}=0.9$ ({\it solid} line) runs. {\it Right
    panel:} The same as the left panel but showing the (3D)
  galaxy--mass correlation function. The error bars plotted
  are 1 $\sigma$ jackknife resampling errors.  For clarity, error bars
  in the simulation results are plotted only for scales at which there
  appears to be some discrepancy between simulations and observations.
  For illustration purposes, the results for the most luminous bin are
  shifted upward by a factor of 10 in both panels.  Also shown on the
  right panel with thin lines are the results when no scatter between
  the luminosity and $V_{\rm max}$ is introduced. The effect of
  scatter on $\Delta \Sigma$ is similar.
\label{fig:comp1}}
\end{figure*}

The effect of simulation box size can be seen in the upper left panel
of the figure.  Each correlation function is shown here for three
different box sizes, all with exactly the same cosmological model and
power spectrum normalization.  Previous studies based on the
mass--mass correlation function have found that the size of the
simulation box affects the amplitude on large scales
\citep[e.g.,][]{itoh_etal92, colombi_etal94,colin_etal99}.  According
to these studies, a small box leads to an underestimate of the
correlation function on scales larger than $\sim 10\%$ of the box
size.  The finite size of the box sets an upper limit on the longest
wavelength of fluctuations that can be present in the simulation. At a
given separation, however, the two-point correlation function, takes
(weighted) contributions from all modes.  This may be the case with
$\Lambda$CDM$_{60c}$ and $\Lambda$CDM$_{80b}$.  From our comparison it
becomes clear that cosmic variance can significantly alter all these
expected trends.  Furthermore, while in the linear regime each mode
evolves independently, in the nonlinear regime mode coupling occurs
and power leaks from large to small scales.  We check the effects of
non-linearity by using the model developed by \cite{smith_etal03} for
the calculation of non-linear mass power spectra. The prediction of
their model is presented as the long dashed-dotted line --- hardly
discernible in the figure, since the agreement of our simulations with
the model is very good.

The upper right panel demonstrates the amount of cosmic variance for
the box sizes we use. Simulations $\Lambda$CDM$_{60b}$ and
$\Lambda$CDM$_{60c}$ are identical in every respect, except that
different random realizations of the initial conditions were used.
The comparisons show that the jackknife error may underestimate the
cosmic variance somewhat.  The effect of varying the value of
$\sigma_8$ is shown in the lower panel for the two 80 $\hMpc$ boxes.
Note that these two simulations used the same random realization of
the initial conditions. Since  the same
structures form in both $80h^{-1}\ \rm Mpc$ simulations, all differences
between the two simulations are thus due to the difference in  $\sigma_{8}$.

Keeping in mind these points, in what follows we focus on two of our
largest volume simulations, $\Lambda$CDM$_{80b}$ and
$\Lambda$CDM$_{120}$. The latter provides more robust results at high
luminosities ($M_{r} \le -21$) due to better statistics than the
smaller volume simulations. Based on the presented box comparisons,
the jackknife error bars should be indicative of how different our
results can be if another simulation were used.

\section{Results}
\label{sec:results}

\subsection{Weak lensing observables}
\label{sec:sdss}

We now compare our simulation results to those recently measured by
S04 using the SDSS.  Using the shapes of the background galaxies,
they estimated the tangential shear $\gamma_{\rm t}$ azimuthally
averaged over thin projected radial annuli from the lens galaxy. The
average tangential shear is given in terms of the quantities related
to the lens mass surface density \citep{kaiser84,miralda-escude91}.
More specifically,
\begin{equation}
\gamma_{t}=\Delta \Sigma / \Sigma_{crit},\ \ \  \Delta \Sigma = \bar{\Sigma} (\le R)-\bar{\Sigma} (R) \,
\label{eq:ds}
\end{equation} 
where $\bar{\Sigma} (\le R)$ is the mean surface density within the
projected radius $R$, $\bar{\Sigma}(R)$ is the azimuthally averaged
surface density at $R$, and $\Sigma_{\rm crit}$ is the critical density
for lensing which depends on the angular diameter distances of the
lens and the source.  To achieve high signal-to-noise S04 stack the
lenses. It is thus desirable to remove
the redshift dependence of the signal, which comes in through
$\Sigma_{\rm crit}$. This dependence is removed if one considers
$\Delta \Sigma = \gamma_{\rm t} \times \Sigma_{\rm crit}$. This is the
quantity that S04 measure.

S04 deprojected their  $\Delta \Sigma$ to obtain the actual 3D galaxy--mass
correlation function. The deprojection is done via an Abel inversion:
\begin{equation}
\xi_{\rm gm}(r)=\frac{1}{\pi \bar{\rho}} \int_{r}^{\infty} \frac{dR}{\sqrt{R^{2}-r^{2}}} 
\left( \frac{d\Delta \Sigma}{dR} + 2\frac{\Delta \Sigma}{R} \right) \ ,
\label{eq:inversion}
\end{equation}
where $r$ is the 3D separation, $R$ the projected separation, and
$\bar{\rho}$ the mean density of the universe. S04 assumed for the present-day matter
$\Omega_{0}=0.27$ 
to obtain the mean density  $\bar{\rho}$ appearing in the inversion. As discussed, 
all of our simulations use $\Omega_{0}=0.3$. Thus when comparing to S04 3D results we show
their results rescaled by a factor of $0.27/0.3$. 

We calculate $\Delta\Sigma$ in simulations by selecting objects from
the entire halo catalog of a certain simulation in accordance with the
magnitude distribution of S04 in a given band (see their Figure 5).
The contribution of each object to the correlation function is
weighted similarly to the weighting of S04. The weight for each lens
depends on $\Sigma_{\rm crit}$, and thus the redshift of the lens, and
on the number of background galaxies available for each lens.  For a
flux-limited sample, this weight as a function of redshift translates
into an effective weight as a function of $r$-band luminosity; to
mimic this selection we thus apply the effective weight for the S04
sample to our simulated catalogs, using a function supplied by
E. Sheldon.

We first compare our results to the direct observable, $\Delta
\Sigma$, which is commonly referred to as the term (projected)
correlation function.  In the case where the 3D correlation function
is a power law, $\Delta \Sigma$ is the projected correlation function
modulo a factor which depends on the slope of the 3D power law and the
average matter density. We also calculate the 3D galaxy--mass
correlation function. In simulations this function can be measured
directly rather than through inversion of $\Delta \Sigma$ by using
Eq.~(\ref{eq:inversion}).

\subsubsection{Luminosity dependence}
\label{sec:lum}

We compare the projected  and the 3D correlation functions,
$\Delta \Sigma$ and $\xi_{\rm gm}$,  measured in our simulations to the
measurements of S04 in the left and right panels of
Figure~\ref{fig:comp1}, respectively.  Note that our faintest bin is
not exactly the same as that of S04, but the faintest galaxies that
are missing from our sample have only a small contribution to the
total signal (E. Sheldon, private communication).

In agreement with S04, we find that the amplitude of the correlation
function increases with luminosity on intermediate ($0.1 - 1\hMpc$)
scales, while it is nearly independent of luminosity on larger scales.
This is indicative of the increase of the effective slope with
luminosity.  In the low luminosity bin our agreement is better for
$\xi_{\rm gm}$ than for $\Delta \Sigma$. As shown by Eq.~(\ref{eq:inversion})  
the value of the 3D
correlation function at given $r$ \ depends on the behavior of $\Delta \Sigma$
and of its derivative  on all scales above $R=r$ (but with contributions
closer to $R=r$ more heavily weighted). 
\begin{figure}[bt]
\vspace{-0.5cm}
\centerline{\epsfxsize=3.5truein\epsffile{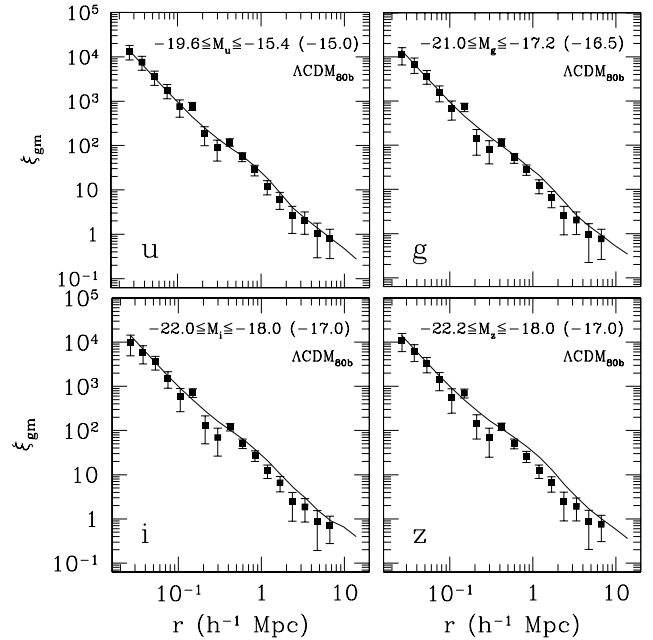}}
\caption{Comparison of the simulation galaxy--mass correlation function 
  ({\it lines}) with the  observations of \cite{sheldon_etal04}  
  ({\it points}) in the $u$, $g$, $i$, and $z$ SDSS bands for
  the fainter sample.  The numbers in parentheses denote the actual
  faint magnitude limit of the \cite{sheldon_etal04} sample in the cases
  where it does not coincide with the faintest magnitude for which
  objects are resolved in the simulations.
\label{fig:other_bands}}
\end{figure}

As discussed in \S \ref{sec:samples}, we introduce an observationally
motivated  scatter in the luminosity--maximum velocity
relation obtained by matching the luminosity and velocity functions.
In the right panel of Figure \ref{fig:comp1} we plot the results
obtained when no scatter between luminosity and velocity is used (thin
lines). The net effect of the scatter is to reduce the amplitude
of the correlation function by a factor of $\sim 2-3$ in the high
luminosity bin, whereas the results for the low luminosity bin are
essentially the same, either with or without scatter. This is expected
given the flatness of the luminosity function at low luminosities, and
its steep fall at high luminosities.  The behavior for $\Delta \Sigma$
is similar. In what follows, only results obtained with scatter are
shown.

The left panel of Figure~\ref{fig:comp1} also shows the difference in
$\Delta\Sigma(R)$ for two simulations with different values of
$\sigma_8$.  Although in \S~\ref{sec:boxes} we showed that this effect
is smaller than the cosmic variance for boxes of this volume, the two
simulations compared here can estimate this effect accurately as they
used the same random realization of the initial conditions and thus
are not subject to differences due to cosmic variance.

We repeated the analysis in the $r$-band for the other SDSS bands and
show the corresponding $\xi_{\rm gm}$ comparisons for the $u$, $g$,
$i$ and $z$ bands in Figure~\ref{fig:other_bands} for the faintest S04
sample.  The luminosities in other bands were assigned using the
observed density-color correlation for the SDSS galaxies (see
\S~\ref{sec:samples}).  The figure shows that the overall agreement of
our simple luminosity scheme, even in the bluer bandpasses, is
surprisingly good. The agreement for the intermediate luminosity bin
of each band is similar.

\subsubsection{Dependence on color}
\label{sec:color_env}

\begin{figure}[bt]
\centerline{\epsfxsize=4truein\epsffile{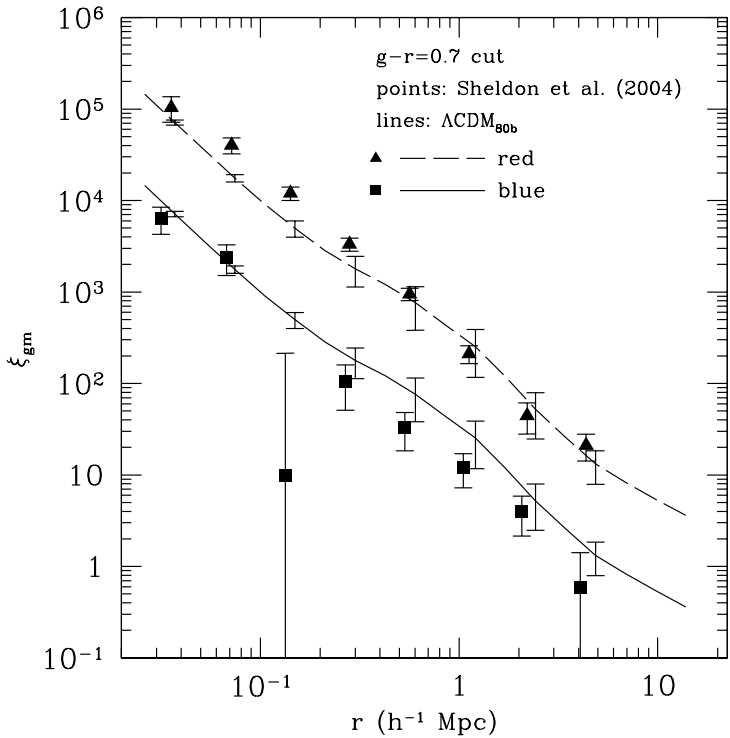}}
\caption{The simulation galaxy--mass correlation function for red 
  ($g-r>0.7$, {\it dashed line}) and blue ($g-r<0.7$, {\it solid
    line}) objects. The corresponding observational results of
  \cite{sheldon_etal04} are also shown by {\it triangles} and {\it
    squares}. For clarity, the results for the red objects are
  displaced upward by a factor of 10. \label{fig:gr}}
\end{figure}

As described in \S~\ref{sec:samples}, we have assigned colors to our
halos based on observational correlation of color and galaxy density.
In Figure~\ref{fig:gr}, we compare the correlation functions of red
and blue galaxies separately to those measured by S04.  In each case,
the color separation is made at $g-r=0.7$, which produces two
subsamples with roughly comparable numbers of objects (as does the
same selection in the data).  The agreement with the data is quite
good in both subsamples, especially if we take into account the fact
that the simulated sample is missing both the dimmest and the most
luminous objects of the S04 total sample, and that cosmic variance is
still an effect given the volumes probed here (note that we do not
plot results from the larger box here because it is not as complete at
the low luminosity end as the simulation shown, 
but the agreement with the larger box is even better).
The agreement indicates that the simple galaxy density-based color
assignment in our simulations is sufficient to explain the
galaxy--mass correlations in the data.

\subsubsection{The $M_{\Delta \Sigma}$-L scaling}
\label{sec:ml}

To quantify the $\Delta \Sigma$-luminosity scaling, many observational
studies use the ``aperture mass'', $M_{\Delta \Sigma}$
\citep{fischer_etal00,mckay_etal01,smith_etal01}.  This is defined as
the quantity obtained by two-dimensional integration of the $\Delta
\Sigma(R)$ up to a certain $R$. Following previous studies, we adopt
the outer scale of $R=260h^{-1}\ \rm kpc$ (physical).  In
observational studies this scale has been chosen because it is  small
enough to avoid a strong contribution from the local environment of
the lenses and large enough to measure a significant signal.

\begin{figure}[t]
\centerline{\epsfxsize=3.9truein\epsffile{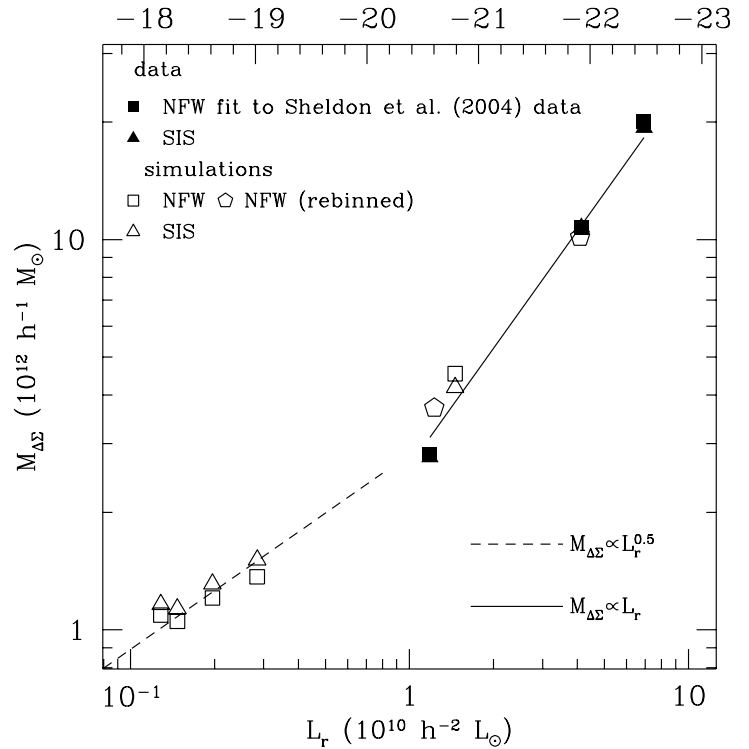}}
\caption{The ``aperture mass'', $M_{\Delta \Sigma}$, as a function of
  $r$-band luminosity. {\it Solid squares} and {\it triangles}
  correspond to the $M_{\Delta \Sigma}$ obtained by fitting the
  \cite{sheldon_etal04} data with a NFW and a SIS profile,
  respectively.  {\it Open squares} and {\it triangles} correspond to
  results obtained from simulation when fitting these profiles. For a
  more direct comparison to the results obtained from the S04 data,
  the pentagons show the simulation results re-binned to match closely
  the S04 luminosity bins.  All results are derived by using
  simulation $\Lambda$CDM$_{\rm 80b}$, with the exception of the
  highest luminosity bin for which the $\Lambda$CDM$_{\rm 120}$ run is
  used to improve the statistics.
\label{fig:m260}}
\end{figure}

The aperture mass simulation results derived in the $r$-band are shown
in Figure ~\ref{fig:m260}.  The data is divided into five bins with
comparable number of objects each.  For each luminosity bin, we fit
the Singular Isothermal Sphere (SIS) and the Navarro, Frenk and White
~\citep[][hereafter NFW]{nfw96,nfw97} profiles to the $\Delta
\Sigma(R)$ in simulations and S04 observations.  The selection and
weighting of our objects is the same as in S04, but the binning
differs: S04 binned their galaxies based on the signal-to-noise ratio,
whereas our bins are chosen to have an equal number of objects per
bin. For a more direct comparison we also re-bin our objects to match
the two lower luminosity bins of S04 (pentagons). 

In all cases the NFW and SIS fits give comparable $M_{\Delta \Sigma}$.
For the SIS fit, $\Delta \Sigma (R) = \Sigma (R)$ (up to an arbitrary
constant), but this is not true for the NFW fit. Note that $M_{\Delta
  \Sigma}$ here refers to the integral of $\Delta \Sigma$ for both
fits. For the SIS profile, $M_{\Delta \Sigma}$ is the true projected
mass (minus a mass sheet), whereas the interpretation for the NFW
profile in terms of an actual mass is more complicated, and one should
perceive $M_{\Delta \Sigma}$ simply as the normalization of $\Delta
\Sigma$ rather than a true projected mass, a point emphasized by
\cite{guzik_seljak01}. A simple interpretation of $M_{\Delta \Sigma}$
as projected mass becomes even more misleading if we take into account
that for many galaxy-size halos $260 \hkpc$ is larger than the virial
radius.

Figure~\ref{fig:m260} shows a good overall agreement in the 
$M_{\Delta\Sigma}--L_r$ relation with
observations.  
Differences between the aperture masses of simulations and observations 
are tracked back to the bins not being exactly the same in the two cases.
More specifically,  the 
 $M_{\Delta\Sigma}$  measured
in simulations is somewhat larger than
observed  for the lowest luminosity bin.
This is due to the difference in the faint limits of our and S04
samples: $M_r=-18$ and $M_r=-17$, respectively:  the contribution from
galaxies with $-17<M_r<-18$ decreases the
$M_{\Delta\Sigma}$ and the amplitude of $\Delta\Sigma(R)$ somewhat, as 
previously mentioned. 

\begin{figure}[tb]
\vspace{-0.5cm}
\centerline{\epsfxsize=3.9truein\epsffile{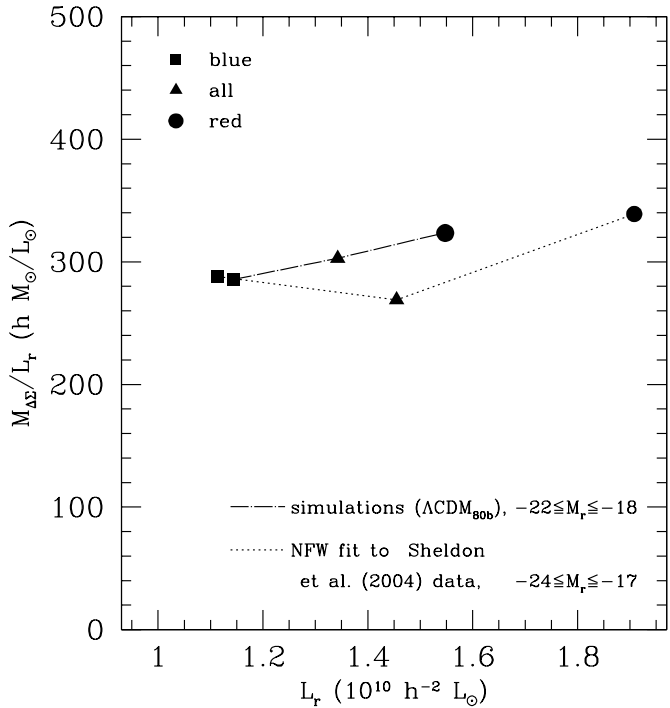}}
\caption{The ``aperture mass''-to-light ratio, $M_{\Delta
  \Sigma}/L_r$, for the entire sample ({\it triangles}), for blue
  ({\it squares}) and red ({\it circles}) galaxies vs average
  luminosity of each sample.  Simulation results are connected by the
  {\it dashed-dotted line}, while the values obtained by fitting the
  $\Delta \Sigma$ measured by \cite{sheldon_etal04} with the NFW
  profile are connected by the {\it dotted line}.  Slightly different
  sample definitions are responsible for offsets in the x-axis.
\label{fig:m260_cuts}}
\vspace{-0.5cm}
\end{figure}

The simulations reproduce the $M_{\Delta\Sigma} \propto L_{r}$ scaling
of the aperture mass with luminosity found by \citet{mckay_etal01}
reasonably well.  Interestingly, our results suggest that the
correlation function flattens significantly for $L_r\lesssim
10^{10}h^{-2}\ \rm L_{\odot}$.  This is in agreement with
\cite{mckay_etal02}, who found that most of the variation of mass with
luminosity is seen for $L_{r} \gtrsim 1.5 \times 10^{10} h^{-2}
L_{\odot}$ galaxies. As is discussed below, such a shallowing is also
observed for the host halo mass-to-light ratio
\citep{vandenbosch_etal03,vale_ostriker04}.

Using this aperture mass measurement, \cite{mckay_etal01} calculated
mass-to-light ratios for the entire sample and for samples separated
by galaxy type, and found that the $M_{\Delta \Sigma}/L$ ratio in the
red wavebands is insensitive to morphology.  In Figure
\ref{fig:m260_cuts} we compare the $M_{\Delta \Sigma}/L_r$ calculated
for our sample with that obtained by fitting a NFW profile to the
S04 sample (this corresponds to the masses designated by solid squares
in Figure~\ref{fig:m260}).  In addition to the value for the entire
sample, we also show values for the blue and red subsamples defined
with the color cut of $g-r=0.7$ (see \S~\ref{sec:samples} and
\S~\ref{sec:color_env}).  The figure shows that the $M_{\Delta
  \Sigma}/L_r$ ratio is fairly constant and thus 
insensitive to the color of the galaxies.  The
average luminosities of the red and blue simulation subsamples differ
from that of the observational subsamples since in simulations we miss
both the dimmest and the most luminous objects.  Note that for the
data of S04 we find approximately $50\%$ higher values of these ratios
than were found by \cite{mckay_etal01} for their sample. Our
results differ from S04 less than the aperture mass-to-light
ratio for their blue and red subsamples differ.

\begin{figure}[t]
\centerline{\epsfxsize=3.5truein\epsffile{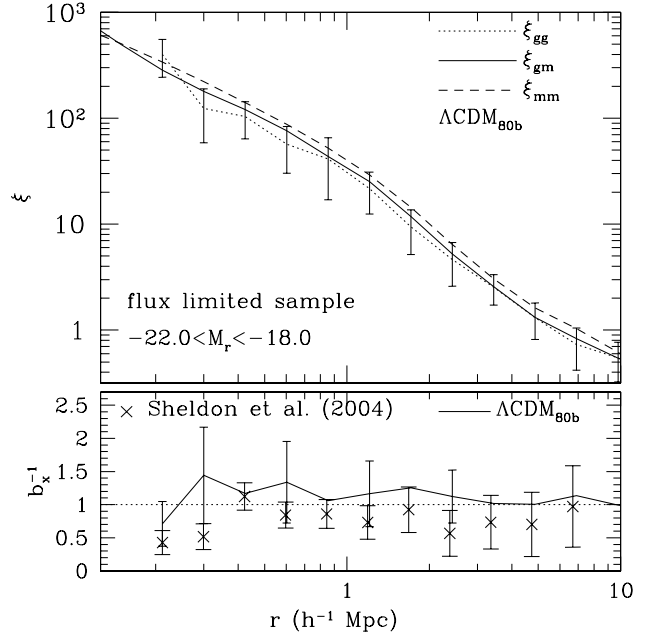}}
\caption{ {\it Top panel:} The galaxy--galaxy, the galaxy--mass and the
  mass--mass correlation functions. For the galaxy--galaxy and the galaxy--mass
  correlations galaxies are selected according to the SDSS selection
  function.  Only objects with $-22<M_{r}<-18$ are used, close to
  the flux limited samples of \cite{sheldon_etal04} and
  \cite{zehavi_etal03}.  {\it Bottom panel:} Inverse cross-bias,
  $b_{x}^{-1}$, as estimated by \cite{sheldon_etal04} ({\it points})
  and as measured in our simulations ({\it solid line}). For clarity,
  error bars in the simulation results are plotted only for scales at
  which there appears to be some discrepancy between simulations and
  observations.
\label{fig:bias}}
\end{figure}

\subsection{Bias}
\label{sec:cf}

The galaxy bias is a measure of how well galaxies trace the underlying
mass distribution.  It is often characterized by the bias parameter,
$b$, defined in terms of the galaxy--galaxy and the mass--mass
correlation functions \citep{kaiser84}

\begin{equation}
b^{2}=\frac{\xi_{\rm gg}}{\xi_{\rm mm}}.
\label{bias}
\end{equation}
The correlation coefficient, $r$, relates the galaxy--mass correlation
function to the mass--mass and galaxy--galaxy correlation functions 
\citep{pen98}:
\begin{equation}
r=\frac{\xi_{\rm gm}}{({\xi_{\rm mm} \xi_{\rm gg}})^{1/2}}.
\label{correl}
\end{equation}

The mass--mass correlation function is not directly measurable, but
weak lensing observations of the galaxy--mass correlations can be
combined with measurements of the galaxy--galaxy correlation function
to estimate the cross-bias parameter,
\begin{equation}
b_{x}=\frac{\xi_{\rm gg}}{\xi_{\rm gm}}=\frac{b}{r}.
\label{crossbias}
\end{equation}
 
Figure \ref{fig:bias} compares the inverse cross bias, $b_{x}^{-1}$,
measured from simulations with that measured by S04 for a
flux-limited sample.  S04 use the inverse galaxy--mass bias because
their $\xi_{\rm gm}$ is much noisier than $\xi_{\rm gg}$. Their flux-
limited sample, used to obtain $\xi_{\rm gm}$, spans the $r$-band
magnitude range $-24 < M_{r} < -17$. For $\xi_{\rm gg}$ they use the
results of \cite{zehavi_etal03} for a flux-limited sample with
$r$-band magnitudes in the range $-22.2 < M_{r} < -18.9$ and selection
function similar to that of S04. S04 corrected for the difference in
average luminosity of the two samples by rescaling the correlation
length of \cite{zehavi_etal03} using its scaling with luminosity in
2dF \citep{norberg_etal01}. For the simulation results we use objects
in the range $-22< M_{r}< -18$ for both $\xi_{\rm gm}$ and
$\xi_{\rm gg}$ and select objects to mimic the selection function of
S04.  The figure shows that the cross-bias in simulations is in very
good agreement with observations. Note that, again, we are missing 
the dimmest and the brightest of the objects used by S04. The cross-bias 
is approximately
scale-independent with $b_x\approx 1$.

\begin{figure}[tb]
\vspace{-0.5cm}
\centerline{\epsfxsize=4truein\epsffile{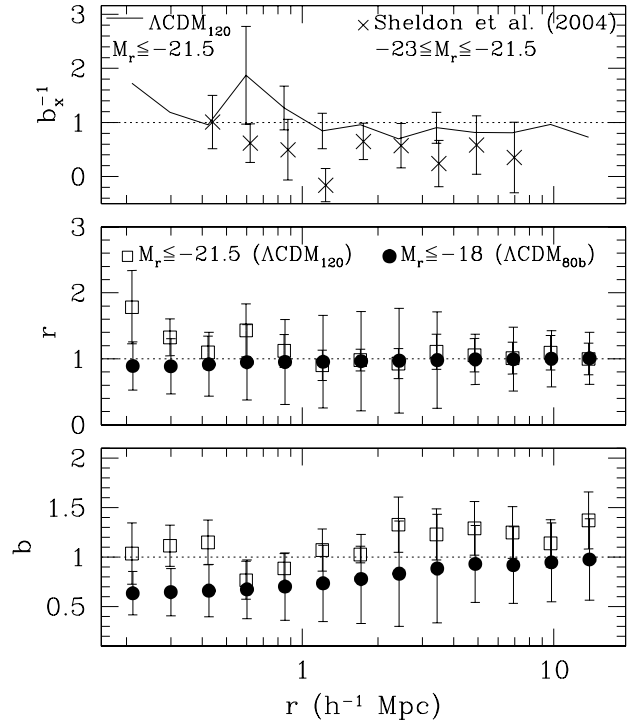}}
\caption{Bias for the volume-limited samples. {\it Top panel:}  
  Inverse cross-bias, $b_{x}^{-1}$, as estimated by
  \cite{sheldon_etal04} ({\it crosses}) for a sample with $-23\le
  M_{r}\le -21.5$ and as measured in our simulations ({\it solid
    line}) for objects with $M_{r}\le -21.5$.  For clarity, error bars
  in the simulation results are plotted only for scales at which there
  appears to be some discrepancy between simulations and observations.
  {\it Middle panel:} Correlation coefficient as a function of scale
  for two volume-limited simulation samples with $M_{r}\le -18$
  ({\it solid circles}) and $M_{r}\le -21.5$ ({\it open squares}).  {\it Bottom
    panel:} Bias as a function of scale for the same samples as
  in the middle panel.
\label{fig:bias_volume}}
\end{figure}

Figure \ref{fig:bias_volume} shows the correlation coefficient $r$ and
bias $b$ for two volume-limited simulation samples of different
luminosities.  In the top panel we also show comparisons of the
cross-bias, $b_x^{-1}$, for the volume-limited sample of S04.  The
simulation results agree with observations within the error bars.  A
small offset of the simulation results toward smaller values of
$b_{x}$ may be due to objects with $-23.0<M_{r}<-22.5$, which are not
present in our simulations due to the small box size and which could
enhance the clustering signal somewhat.

Semi-analytic galaxy formation models suggest that $b_{x}$ is a direct
measure of the bias $b$ on large scales, or, equivalently, that the
correlation coefficient $r$ approaches unity for $r\gtrsim 1 \hMpc$
\citep{guzik_seljak01,berlind_weinberg02}.   Figure
\ref{fig:bias_volume} shows that the correlation coefficient is indeed
approximately unity on scales $\geq 1 \hMpc$ in our simulations.  This
therefore confirms that on these scales the cross bias is expected to
be a fair measure of the standard bias $b$ for both bright and faint
samples.

The scale dependence of bias measured here is in agreement with
previous studies \citep{colin_etal99}. Note that the lower luminosity
sample exhibits an ``antibias'' at $r\lesssim 3h^{-1}\ \rm Mpc$.  For
the high-luminosity sample the shape of the bias profile is similar to
that of the low-luminosity sample: it is scale-independent 
on large scales and decreases on smaller scales. The decrease
on small scales has been attributed to dynamical friction and tidal
destruction processes in the high density regions of clusters and
groups \citep{kravtsov_klypin99,zentner_etal04}. An increase at small
separations, $r\lesssim 0.5h^{-1}\ \rm Mpc$, is likely due to the
tendency of the brightest galaxies to lie near the center of groups
and clusters \citep{seljak00,weinberg_etal04}. For the same reason the
correlation coefficient of the bright subsample increases above unity  for
separations  $r\lesssim 1 h^{-1} \rm Mpc$.

\section{Discussion}
\label{sec:discussion}
\subsection{Luminosity dependence of galaxy--mass correlations}
\label{sec:steepening}

\begin{figure}[t]
\vspace{-0.5cm}
\centerline{\epsfxsize=3.9truein\epsffile{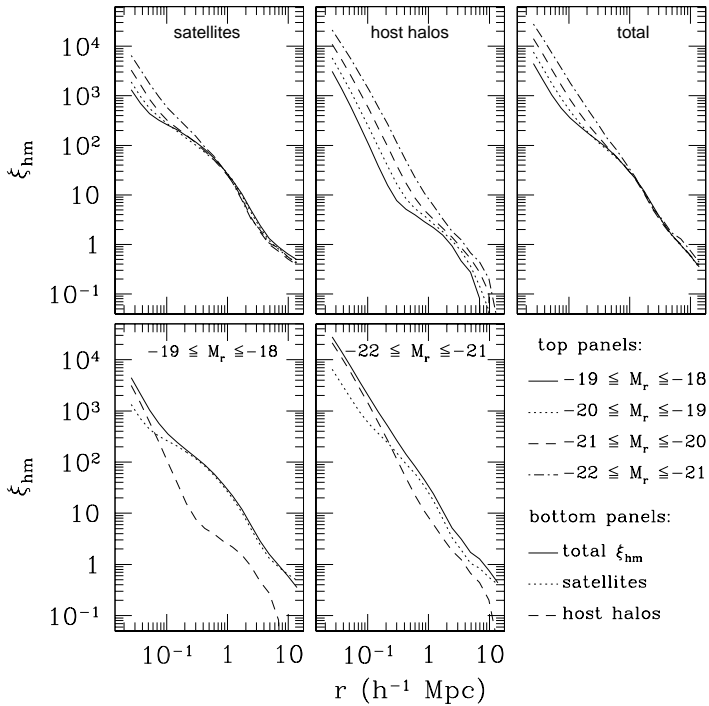}}
\caption{{\it Top row of panels:\/} Volume-limited 
  halo--mass correlation function for four different $r$-band
  luminosity bins and contributions to the correlation function from
  satellite and host halos. {\it Upper left panel:} The satellite
  contribution.  {\it Middle upper panel:} The contribution of host
  halos.  {\it Right upper panel:} total correlation function.  {\it
    Lower left panel:} The (total) halo--mass correlation function
  ({\it solid line}) for the $-19\geq M_r\geq -18$ luminosity bin, the
  satellite contribution ({\it dotted line}), and the host halo
  contribution ({\it dashed line}) for the dimmest luminosity bin.
  {\it Lower right panel:} The same as in the lower left panel but for
  the brightest luminosity bin.
\label{fig:steepening}}
\vspace{-0.5cm}
\end{figure}
\begin{figure}[th]
\centerline{\epsfxsize3.5truein\epsffile{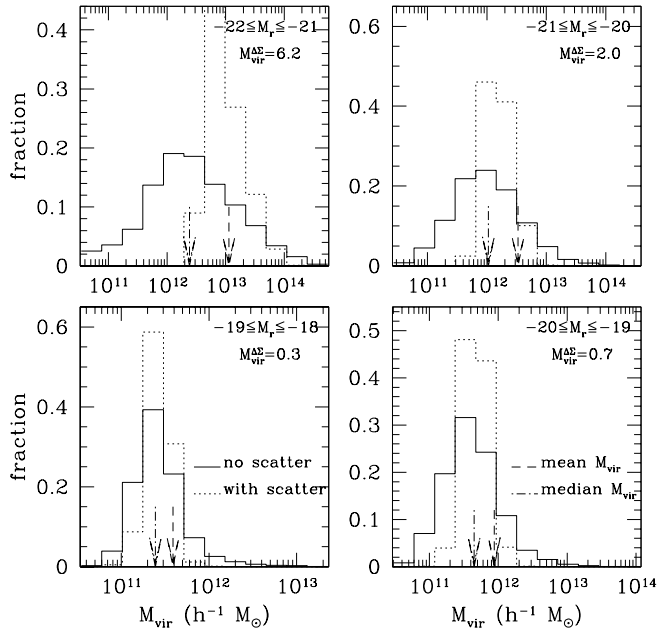}}
\caption{Halo virial mass distributions (defined using an overdensity
  of 180 with respect to the mean density of the universe) for central
  galaxies in four luminosity bins with and without scatter in the
  $V_{max}$-luminosity relation ({\it solid} and {\it dotted}
  histograms, respectively). $M_{\rm vir}^{\Delta \Sigma}$ is the
  virial mass obtained by fitting the $\Delta \Sigma$ of the
  corresponding subsample of galaxies, in units of $10^{12} h^{-1}
  M_{\odot}$.  For the distributions with scatter, the mean and median
  virial masses are indicated with {\it dashed} and {\it dot-dashed}
  arrows, respectively.
\label{fig:mass_distr}}
\end{figure}

In the previous section we showed that simulations can reproduce the
luminosity dependence of the galaxy--mass correlation function
observed for the S04 sample, if a reasonable amount of scatter between
$V_{\rm max}$ and luminosity is assumed. The luminosity dependence of
these correlations is weak at large scales ($\gtrsim 1 \hMpc$) and
strong at small scales ($0.1<R<1\hMpc$).  To understand the origin of
this behavior, we compute contributions to the correlation function
$\xi_{\rm hm}$ from host halos (``central galaxies'') and from halos
classified as satellites (see \S~\ref{sec:haloid}).
Figure~\ref{fig:steepening} shows the results for the simplest case of
the volume-limited samples with $r$-band luminosities in four
different bins, $[-19,-18]$, $[-20,-19]$, $[-21,-20]$, and
$[-22,-21]$.  We adopt the convention of including the $-1$ term
appearing in the definition of the total correlation function in the
host halo term.

The bottom panels of Figure~\ref{fig:steepening} show that the
relative contribution of the host halos compared to that of satellites
is considerably larger at high luminosities, as expected.  The central
galaxy contribution at scales $r\lesssim 1h^{-1}\ \rm Mpc$ changes
with luminosity only in amplitude, not in shape.  At the high
luminosity end, most galaxies are central galaxies and the increase of
the correlation amplitude with luminosity reflects the increase in
average mass and density of their host halos
\citep{berlind_weinberg02}.  At low luminosities, and at
intermediate--large separations, the satellite contribution is
significant and leads to the development of a 'shoulder' in the
overall correlation function, since outside the typical satellite
radius, one starts to measure the density of the environment in which
the satellites are embedded.  The existence of this feature makes the
total correlation function at relatively low luminosities effectively
shallower than that at high luminosities.

\subsection{Average halo masses and density profiles\\ from weak lensing} 
\label{sec:masses}

An important question about weak lensing observations is whether
they will allow us to extract information about the average halo mass
and mass profile of the lenses.  Stacking the lenses
to achieve high signal-to-noise makes the interpretation of the
average mass difficult. It has been noted \citep[see,
e.g.,][]{guzik_seljak01, yang_etal03} that complications arise for
several reasons including the scatter in sizes and masses of objects
and the fact that a fraction of galaxies may be
located within a group or cluster of galaxies.

These complications could be limited if only the central galaxy of
each halo was used to measure the correlation signal. However, even in
this case the interpretation of mass is not straightforward. In Figure
\ref{fig:mass_distr} we show the distribution of virial
masses\footnote{Recall that the virial mass is defined as the mass
  within the radius at which an overdensity of 180 times the mean
  matter density of the universe is reached.} for central galaxies in
four luminosity bins.  The distributions in cases with and without
scatter are shown. The distribution of masses in the case without
scatter has a finite width because there is scatter between virial
mass and $V_{\rm max}$. The figure shows that introducing scatter
between $V_{\rm max}$ and $L_r$ significantly broadens the mass
distributions and makes them non-gaussian. The broadening is largest
in the most luminous bin. As a result, it becomes difficult to assign
a specific mass to a given (even narrow) luminosity range.  For
example, the figure shows that the mean mass is significantly larger
than the median for every luminosity bin.  The best fit virial mass,
$M_{\rm vir}^{\Delta \Sigma}$, obtained by fitting $\Delta \Sigma$ for
the corresponding galaxy sample to a NFW profile quoted in the figure
is a different effective averaging, and gives values between the
median and the mean masses.

In agreement with \cite{yang_etal03}, we find a tight correlation
between $M_{\rm vir}^{\Delta \Sigma}$ and the aperture mass $M_{\Delta
  \Sigma}$. Both quantities are derived from the same fit, and thus a
correlation is expected, but without information on the halo
concentrations, one cannot be derived from the other. Nevertheless, as
\cite{yang_etal03} showed, this correlation is recovered easily if one
assumes the well established correlation between the virial mass and
concentration \citep{bullock_etal01a}, which in the absence of scatter
between concentration and virial mass renders the NFW effectively a
one parameter fit. Given the averaging  done to obtain $\Delta
\Sigma$, the average concentration-mass correlation is sufficient.  In
agreement with \cite{yang_etal03}, we find the following scaling for
both central galaxies and galaxies in low density environments
\begin{equation}
M_{\rm vir}^{\Delta \Sigma}= (0.31 \pm 0.01) \times  M_{\Delta \Sigma}.
\end{equation}  
Here both masses are in units of $10^{12} h^{-1} M_{\odot}$.  The
scaling is exactly the same when no scatter is used, 
since this relation reflects the properties of the average mass
density profiles of our dark matter halos.

The existence of significant scatter between luminosity and mass even
for central galaxies indicates that the mass, $M_{\rm vir}^{\Delta
  \Sigma}$, derived from weak lensing cannot be interpreted in a
straightforward way as the mass for galaxies of a given luminosity.
Instead, there is a rather broad distribution of masses. The
mass--luminosity relations should therefore be quoted and interpreted
with caution. Figure~\ref{fig:mass_distr} shows that the amplitude of
the relation will depend on which mass one considers.  The actual
distribution depends on the still uncertain details of the scatter
between mass and light, but even if the true relation is different
from what we have assumed here, it will be incomplete if the
distribution of masses is not specified.  In conclusion, the mass
distribution at fixed luminosity can be wide enough so that the
concept of a mean mass becomes obscure. Referring to $M_{\rm
  vir}^{\Delta \Sigma}$, the mass of $L_{\ast}$ galaxies that we find
is $\approx 2\times 10^{12}h^{-1}\ \rm M_{\odot}$ as shown for the
$-21\leq M_r\leq -20$ bin in Figure~\ref{fig:mass_distr} and is in
general agreement with the mass found by \cite{yang_etal03} and the
best fit values derived by \cite{guzik_seljak02} from the fit to the
SDSS weak lensing data.

A related question is whether a meaningful average density profile of
halos can be recovered from lensing data.  Our results show that here,
again, one has to focus on either the central galaxies or galaxies in
low density environments. As we show in \S \ref{sec:steepening}, for
luminous galaxies the galaxy--mass correlation is dominated by the
contribution of central galaxies, especially at small scales. For
high-luminosity samples it is therefore meaningful to interpret
$\Delta\Sigma(R)$ in terms of the average halo density profile at  
$r\lesssim 100-200h^{-1}$~kpc.

\section{Conclusions}
\label{sec:conclusions}

We have presented a detailed comparison of galaxy--mass correlation
functions measured in high-resolution cosmological simulations of the
concordance $\Lambda$CDM cosmology with the most recent weak lensing
SDSS observations. We found that the simple recipe of assigning a
luminosity to a halo of a certain maximum circular velocity by
matching the subhalo velocity function in simulations to the observed
luminosity function leads to good agreement with the observed
galaxy--mass correlation and its dependence on luminosity, if an
observationally-motivated amount of scatter is introduced in the
$V_{\rm max}-L_{r}$ relation. Our main results and conclusions can be
summarized as follows:

\begin{itemize}
\item[$\bullet$] The simulations reproduce the galaxy--mass
  correlation function measured by \citet{sheldon_etal04} in SDSS and
  the observed dependence of its shape and amplitude on luminosity.
  Interestingly, the agreement for bright samples seems to require
  some scatter between luminosity and circular velocity.  The amount
  of scatter required is consistent with the observed scatter.
  
\item[$\bullet$] The galaxy--mass correlation function for central
  galaxies is a strong function of the galaxy luminosity (halo
  velocity), while $\xi_{\rm gm}$ for satellite galaxies is only
  weakly sensitive to luminosity.  The luminosity dependence of the
  correlation function as a whole is thus determined primarily by the
  increasing contribution of bright central galaxies relative to the
  satellite galaxies at bright luminosities.  Conversely, the
  correlation function gets shallower with decreasing luminosity
  because it is increasingly dominated by the contribution from the
  satellite (non-central) galaxies in halos at intermediate and large scales.
 
\item[$\bullet$] We use the color-density correlation observed in the
  SDSS to assign colors to the galactic halos in simulations and
  compare the simulation results to observations in other SDSS bands,
  from $u$ to $z$. In each band the agreement between galaxy--mass
  correlation function in simulations and observations is remarkably
  good. The simulations also reproduce the observed trend of the
  galaxy--mass correlation function with the $g-r$ color.
  
\item[$\bullet$] In agreement with previous studies, we find that the
  aperture mass-to-light ratio, $M_{\Delta\Sigma}/L_r$, is independent
  of galaxy color. For $L_r\gtrsim 10^{10} h^{-2} L_{\odot}$,
  $M_{\Delta\Sigma}/L_r$ is approximately independent of luminosity.
  For both central galaxies and galaxies in low density environments the
  best fit virial mass $M_{\rm vir}^{\Delta \Sigma}$ correlates tightly with the observed aperture
  mass. The best fit virial mass lies in between the median and the mean actual virial
  mass when scatter between $V_{max}$ and luminosity is used.  
  
\item[$\bullet$] We compare the cross bias, $b_x\equiv b/r$, measured
  in simulations and observations and find a good agreement at all
  probed scales.  The correlation coefficient, $r$,
  obtained for a volume-limited sample similar to that analyzed by
  S04, is close to unity on scales $\gtrsim 1 \hMpc$. This indicates
  that the cross bias measured in weak lensing observations does
  measure the actual bias $b$ of galaxy clustering on these scales.

\item[$\bullet$] We show that for luminous galaxies ($M_r<-21$) the
  galaxy--mass correlation function at $r\lesssim 100-200h^{-1}\ \rm kpc$
  can be interpreted as the average density profile of these galaxies.
  We also show that the masses obtained by 
  the $\Delta\Sigma(R)$ measured in weak lensing
  observations cannot be interpreted in a straightforward way as the 
  mass for galaxies of a given luminosity. In the presence of scatter
  between mass and light, the obtained mass-luminosity relations should be
  quoted and interpreted with caution.   

\end{itemize}    

\acknowledgements 

We are grateful to Erin Sheldon for invaluable discussions and help
understanding the data, for providing his data in electronic format,
and for comments on the manuscript.  We would also like to thank
Anatoly Klypin for running the simulations used in this study, Brandon
Allgood for running some of the halo catalogs used, Michael Blanton
for communicating to us the results on the scatter between $r$- and
$K$-band luminosities for the SDSS galaxies, Arieh Maller for helpful
discussions about scatter and comments on a draft, Chris Miller for
allowing us to use results from the Value Added Catalog he has
compiled for the SDSS DR1, and Uros Seljak and David Weinberg for
useful comments on the manuscript.  This work was supported by the
National Science Foundation (NSF) under grants No.  AST-0206216 and
AST-0239759, by NASA through grants NAG5-13274 and NAG5-12326, and by
the Kavli Institute for Cosmological Physics at the University of
Chicago.  RHW was supported by NASA through a Hubble Fellowship
awarded by the Space Telescope Science Institute, which is operated by
the Association of Universities for Research in Astronomy, Inc, for
NASA, under contract NAS 5-26555.  The simulations used in this study
were performed on the IBM RS/6000 SP3 system at the National Energy
Research Scientific Computing Center (NERSC) and on the Origin2000 at
the National Center for Supercomputing Applications (NCSA).  This
study has used data from the Sloan Digital Sky Survey (SDSS).  Funding
for the creation and distribution of the SDSS Archive has been
provided by the Alfred P. Sloan Foundation, the Participating
Institutions, the National Aeronautics and Space Administration, the
National Science Foundation, the U.S. Department of Energy, the
Japanese Monbukagakusho, and the Max Planck Society. The SDSS Web site
is http://www.sdss.org/.

\bibliography{ms.bbl}

\end{document}